\documentclass[prd,twocolumn,showpacs,amsmath,amssymb]{revtex4-1}
\usepackage{amssymb}
\usepackage{amsfonts}
\usepackage{overpic,graphicx}
\usepackage{textcomp,wasysym}
\usepackage[perpage,symbol]{footmisc}
\usepackage{lineno}
\usepackage{multirow}
\usepackage{xcolor}
\usepackage{units}
\usepackage{bigstrut}
\usepackage{array}
\usepackage{amsmath}

\begin{document}

\title{\boldmath Search for Baryon and Lepton Number Violation in $J/\psi\to\Lambda_c^+e^-+c.c.$}

\author{
\begin{small}
\begin{center}
M.~Ablikim$^{1}$, M.~N.~Achasov$^{9,d}$, S.~Ahmed$^{14}$, M.~Albrecht$^{4}$, A.~Amoroso$^{53A,53C}$, F.~F.~An$^{1}$, Q.~An$^{40,50}$, J.~Z.~Bai$^{1}$, Y.~Bai$^{39}$, O.~Bakina$^{24}$
% ...
\\
\vspace{0.2cm}
(BESIII Collaboration)\\
\vspace{0.2cm} {\it
$^{1}$ Institute of High Energy Physics, Beijing 100049, People's Republic of China\\
% ...
$^{4}$ Bochum Ruhr-University, D-44780 Bochum, Germany\\
% ...
$^{9}$ G.I. Budker Institute of Nuclear Physics SB RAS (BINP), Novosibirsk 630090, Russia\\
% ...
$^{14}$ Helmholtz Institute Mainz, Johann-Joachim-Becher-Weg 45, D-55099 Mainz, Germany\\
% ...
$^{24}$ Joint Institute for Nuclear Research, 141980 Dubna, Moscow region, Russia\\
% ...
$^{39}$ Southeast University, Nanjing 211100, People's Republic of China\\
$^{40}$ State Key Laboratory of Particle Detection and Electronics, Beijing 100049, Hefei 230026, People's Republic of China\\
% ...
$^{50}$ University of Science and Technology of China, Hefei 230026, People's Republic of China\\
% ...
$^{53}$ (A)University of Turin, I-10125, Turin, Italy; (B)University of Eastern Piedmont, I-15121, Alessandria, Italy; (C)INFN, I-10125, Turin, Italy\\
% ...
$^{d}$ Also at the Novosibirsk State University, Novosibirsk, 630090, Russia\\
% ...
}
\end{center}
\vspace{0.4cm}
\end{small}
}

\noaffiliation{}

\begin{abstract}
Using $1.31\times10^9$ $J/\psi$ events collected by the BESIII detector at the Beijing
Electron Positron Collider, we search for the process  $J/\psi\to\Lambda_c^+e^-+c.c.$ for the first time.  In this process, both baryon and lepton number conservation is violated. No signal is found and the upper limit on the branching fraction $\mathcal{B}(J/\psi\to\Lambda_c^+e^-+c.c.)$ is set to be $6.9\times10^{-8}$ at the 90\% Confidence Level.
\end{abstract}


\maketitle

The observed matter--antimatter asymmetry in the universe composes a serious challenge to our understanding of nature. 
The Big Bang theory, the prevailing cosmological model for the
evolution of the universe, predicts exactly equal numbers of baryons
and antibaryons in the dawn epoch. However, the observed baryon number
(BN) exceeds the number of antibaryons by a very large ratio, currently estimated at $10^{9}\sim 10^{10}$~\cite{ref::universe}. To give a reasonable interpretation of the baryon-antibaryon asymmetry, Sakharov proposed three principles~\cite{ref::sakharov}, the first of which is that BN conservation must be violated. 
Many proposals predict BN violation within and beyond the SM. Among them, proposals that
evoke the spontaneous breaking of a large gauge group are especially appealing.
In these models, several heavy gauge bosons emerge whose couplings to matter explicitly violate both baryon and lepton number conservation simultaneously.
The $\rm{SU}(5)$ grand unification theory (GUT)~\cite{ref::su5}, which covers the SM gauge group $\rm {SU}_c(3)\times \rm{SU}_L(2)\times \rm{U}_Y(1)$, was first proposed as a minimal extension of the SM. In that scenario, two extra gauge bosons, $X$ and $Y$, with charges of 4/3 and 1/3, the so-called leptoquarks~\cite{ref::leptoquark}, exist and can violate baryon-lepton number conservation. In the unification picture, their masses are about $10^{15}$~GeV/$c^2$. 
Unfortunately, this simplest $\rm{SU}(5)$ model is ruled out because its prediction of the proton lifetime is several orders of magnitude smaller than the experimental lower limits~\cite{ref::protonlife}.  However, this does not rule out the need to search for grand unification theories that allow for BN violation. For example, the $\rm{SO}(10)$, the $\rm{E}6$ and the flipped $\rm{SU}(5)$ models all predict a longer proton lifetime that is not in conflict with the present data. 

Searches for physics beyond the standard model (`new physics') with
collider experiments are complementary to searches with specifically designed precision
detection experiments.  For example, the existence of dark matter (DM)
is strongly suggested by cosmological observations, and searches for
particle candidates of the dark sector are carried out both at
$e^+e^-$ and $pp$ collider experiments and in dedicated direct
detection experiments.  Similarly, searches for Majorana neutrinos at
charm/bottom factories supplement the neutrino-less double beta decay
experiments. The two independent ways of searching for new physics are
fruitfully supporting each other. Therefore, searching for BN
violation processes at collider experiments might provide different
and complementary information from the proton decay experiments.

\begin{figure}[htbp]
\begin{center}
\begin{minipage}[t]{9cm}
\includegraphics[width=4.0cm]{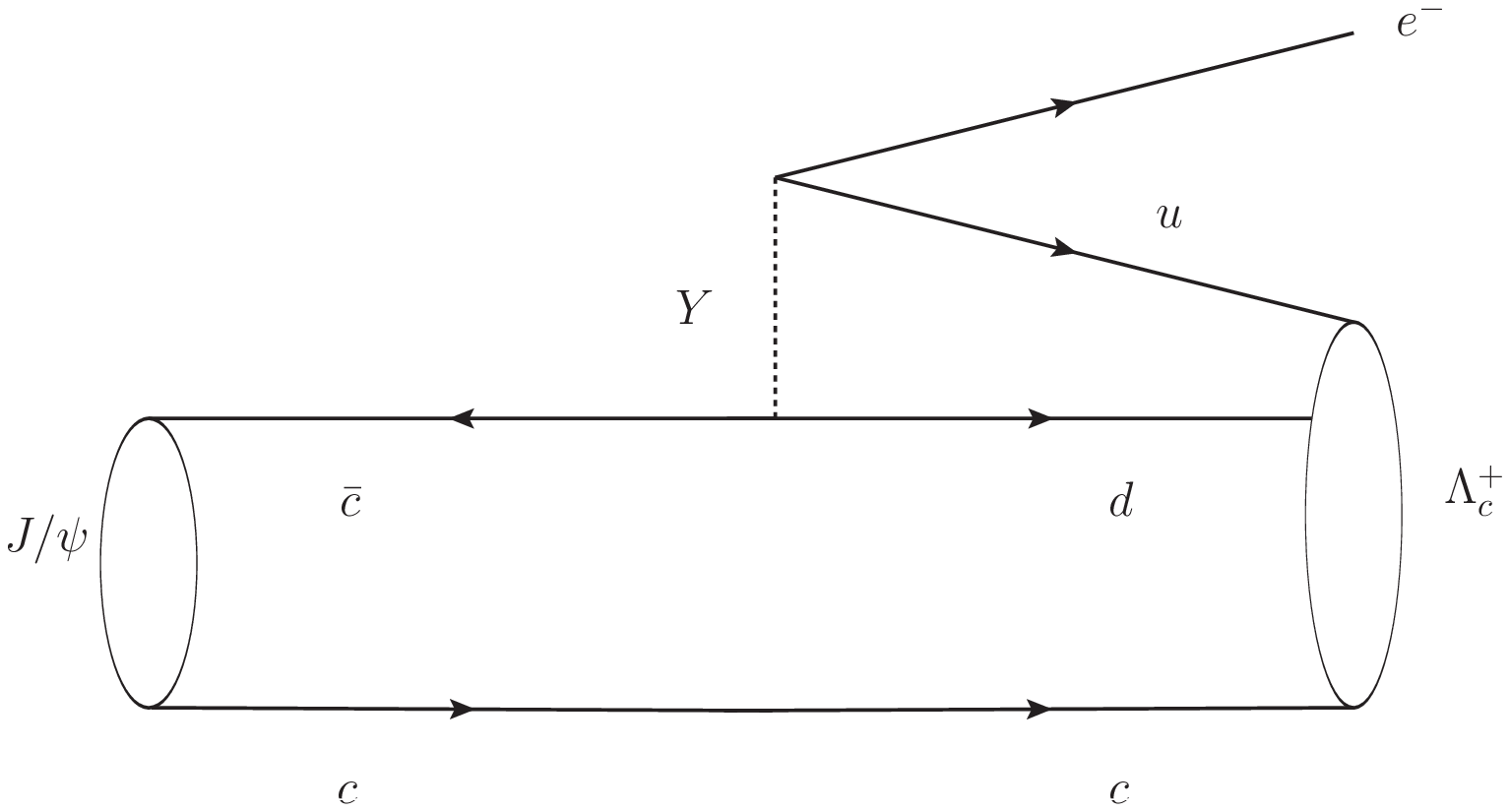}
\hspace{0.1cm}
\includegraphics[width=4.0cm]{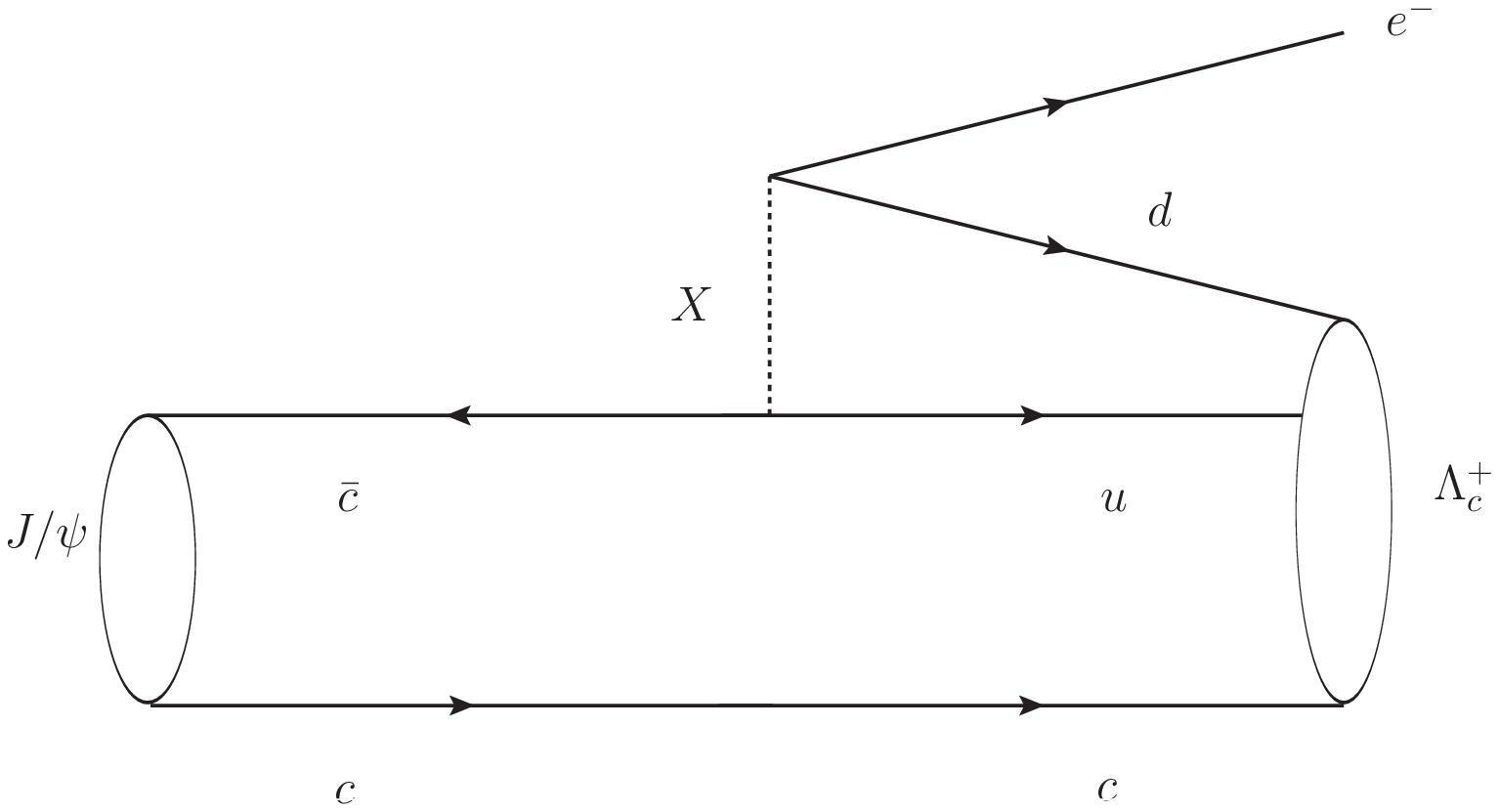}
\end{minipage}
\caption{Decay diagrams for $J/\psi\to\Lambda_c^+e^-$.} \label{fig::gut}
\end{center}
\end{figure}

In any case, the matter--antimatter asymmetry in the universe is an observable fact.  The absence so far of an experimental observation of proton decay, which is predicted by GUT, does not imply by any means that BN is absolutely conserved. Therefore, an alternative approach to test the GUT scheme at collider experiments has been devised.  The CLEO Collaboration searched for very rare processes which violate BN conservation in decays of heavy-flavor mesons.  In particular, they suggested to look for the process $D^0\to \bar p+e^+$, which is an inverse process of $p\to \pi^0 e^+$ at the quark level.  The upper limit from the CLEO measurement is $10^{-5}$~\cite{ref::cleo-d0} at 90\% Confidence Level (CL), limited by low statistics.  Instead, thanks to the huge data sample of $J/\psi$ decays produced at the BESIII experiment, we are able to study the analogous process $J/\psi\to \Lambda_c^+ e^-$, as shown in Fig.~\ref{fig::gut}, with  much higher statistics and therefore better sensitivity.

In this Letter, we analyze the $J/\psi$ data sample collected with the BESIII~\cite{ref::detector} detector operating at the BEPCII storage ring~\cite{ref::collider} to search for the SM forbidden baryon-lepton number violating decay $J/\psi\to \Lambda_c^+e^-$ (charge conjugation is implied throughout this Letter). Based on this analysis, we set an upper bound on the rate of $J/\psi\to \Lambda_c^+e^-$. 

The BESIII detector has a geometric acceptance covering $93$\% of the $4\pi$ solid angle and consists of the following main components.
(1)~A small-celled main drift chamber~(MDC) with $43$ layers is used to track charged particles. The average single-wire resolution is $135$~$\mu$m, the momentum resolution for $1$~GeV/$c$ charged particles in a $1$~T magnetic field is $0.5$\%, and the specific energy loss ($dE/dx$) resolution is better than $6$\%.
(2)~An electromagnetic calorimeter~(EMC) is used to measure photon energies. The EMC is made of $6240$ CsI(Tl) crystals arranged in a cylindrical shape (barrel) plus two endcaps. For $1.0$~GeV photons, the energy resolution is $2.5$\% in the barrel and $5$\% in the endcaps, and the position resolution is $6$~mm for the barrel and $9$~mm for the endcaps.
(3)~A time-of-flight (TOF) system is used for particle identification (PID). It is composed of a barrel made of two layers, each consisting of $88$ pieces of $5$~cm thick and $2.4$~m long plastic scintillators, as well as two endcaps each with 96 fan-shaped $5$~cm thick plastic scintillators. The time resolution is 80~ps in the barrel and $110$~ps in the endcaps, providing a $K/\pi$ separation of more than $2\sigma$ for momenta up to about $1.0$~GeV/$c$.
(4)~A muon chamber system for muon detection is made of resistive plate chambers arranged in $9$ layers in the barrel and $8$ layers in the endcaps and is incorporated into the return iron yoke of the superconducting magnet.

Optimization of the event selection criteria and estimation of physics backgrounds are performed through Monte Carlo (MC) simulations of background and signal samples. The GEANT4-based~\cite{ref::geant4} simulation software BOOST~\cite{ref::boost} includes the geometric and material description of the BESIII detector, the detector response and digitization models, and also keeps track of the detector running conditions and performance. The analysis is performed in the framework of the BESIII Offline Software System (BOSS)~\cite{ref::boss} which takes care of the detector calibration, event reconstruction and data storage.
Inclusive MC events of $J/\psi$ decays are generated by the KKMC~\cite{ref::kkmc} generator around $\sqrt{s}$ = $3.097$ GeV, in which the beam energy and spread are set to the values measured at BEPCII, and initial state radiation (ISR) is considered. The known $J/\psi$ decays are generated by BesEvtGen~\cite{ref::ISR-lundcharm,ref::evtgen} with branching fractions set to the world average values according to the Particle Data Group (PDG)~\cite{ref::pdg2014}, and the remaining unknown decays are modeled by Lundcharm~\cite{ref::ISR-lundcharm}.

%
%
We search for the decay  $J/\psi\to\Lambda_c^+e^-$, where the $\Lambda_c^+$ is reconstructed through the decay $\Lambda_c^+\to p K^-\pi^+$. In each event, at least four charged tracks are required.
All charged tracks are required to satisfy a geometrical acceptance of $|\rm{cos}\theta| < 0.93$, where $\theta$ is the polar angle of the charged track. Each track must originate from the interaction region, defined as $R_{xy} <1.0$ cm and $|R_{z}| < 10.0$ cm, where $R_{xy}$ and $R_{z}$ are the distances of the closest approach to the interaction point of the track in the $xy$-plane and $z$-direction, respectively. Events with exactly four selected charged tracks with zero net charge are retained for further analysis.

%
%
For charged particle identification, we use a combination of the energy loss $dE/dx$ in the MDC, time of flight in the TOF, and the energy and shape of clusters in the EMC to calculate the CL for the electron, pion, kaon, and proton hypotheses ($CL_e$, $CL_{\pi}$, $CL_K$ and $CL_p$).
The electron and positron candidates are required to satisfy $CL_e>0.001$ and $CL_e/(CL_e+CL_K+LC_{\pi})>0.8$. Other charged tracks will be considered a pion, kaon or proton, according to the highest CL of the corresponding hypothesis.

In order to improve the mass resolution, a kinematic fit enforcing energy-momentum conservation is performed. To suppress contamination from other decay modes with four charged tracks, 
six different combinations of mass
assignments are considered:
$p K^-\pi^+e^-$, $\pi^+\pi^-\pi^+\pi^-$, $K^+K^-K^+K^-$, $\pi^+\pi^-K^+K^-$, $\pi^+\pi^-p\bar p$ and $K^+K^-p\bar p$. If the kinematic fit procedure for the $p K^-\pi^+e^-$ mass assignment is successful and the goodness of fit for this hypothesis is the best among these six assignments, then the event is accepted for further analysis.

Based on the MC simulation, the $\Lambda_c^+$ signal window is defined to be $(2.27, 2.30)$ GeV/$c^2$ in the $pK^-\pi^+$ invariant mass distribution. This corresponds to a range of $\pm4$ times the mass resolution around the $\Lambda_c^+$ nominal mass.
The detection efficiency is determined to be $(35.43\pm0.02)\%$
based on simulated $J/\psi\to\Lambda_c^+e^-\to pK^-\pi^+e^-$ events, where the $\Lambda_c^+$ decay is modeled by a dedicated generator according to the result of a Partial Wave Analysis (PWA) of the decay $\Lambda^+_c\to pK^-\pi^+$~\cite{ref::lambdac2pkpi}.
Besides the non-resonant $3$-body decay process, processes with intermediate states (such as $\Delta^{++}$, $\Delta(1600)^{++}$, excited $\Lambda$ states, excited $\Sigma$ states), as well as the corresponding interferences, are also included in the helicity amplitudes. Parity conservation is not required since this is a weak decay.
The data and MC simulation for the decay $\Lambda_c^+\to pK^-\pi^+$ are compared and found to be in good agreement, based on $567$~pb$^{-1}$ of experimental data taken at $\sqrt{s}=4.599$~GeV, just above the threshold for $\Lambda_c$ pair production~\cite{ref::lambdac2pkpi}. This consistency leads to a negligible systematic uncertainty due to the generator.

%
%
The background from $J/\psi$ decays is investigated using an inclusive MC sample which has the same size as the $J/\psi$ data sample. No background events are found in the signal window.
The background from QED processes is studied with other simulated MC samples of $e^+e^-\to q\bar{q}$, $e^+e^-\to(\gamma)e^+e^-$ and $e^+e^-\to(\gamma)\mu^+\mu^-$ which correspond to 40, 1.5 and 30 times the $J/\psi$ data, respectively. Most of these backgrounds are rejected by the PID requirements and the kinematic fit. The normalized number of surviving background events is $0.03$, which is from wrong PID in the process $e^+e^-\to K^+K^-\pi^+\pi^-$. 
The background from QED processes is also verified by using experimental data samples taken away from the $J/\psi$ and $\psi(3686)$ mass regions, including data taken at $3.08$ GeV, $3.65$ GeV, and scan data sets covering the energy range from $2.23$ to $4.59$ GeV. No events are found in the signal window after taking into account the differences in the integrated luminosities, the cross sections, the particle momenta, and the beam energies~\cite{ref::PRL109-042003}.

\begin{figure}[htbp]
\begin{center}
\includegraphics[width=8cm]{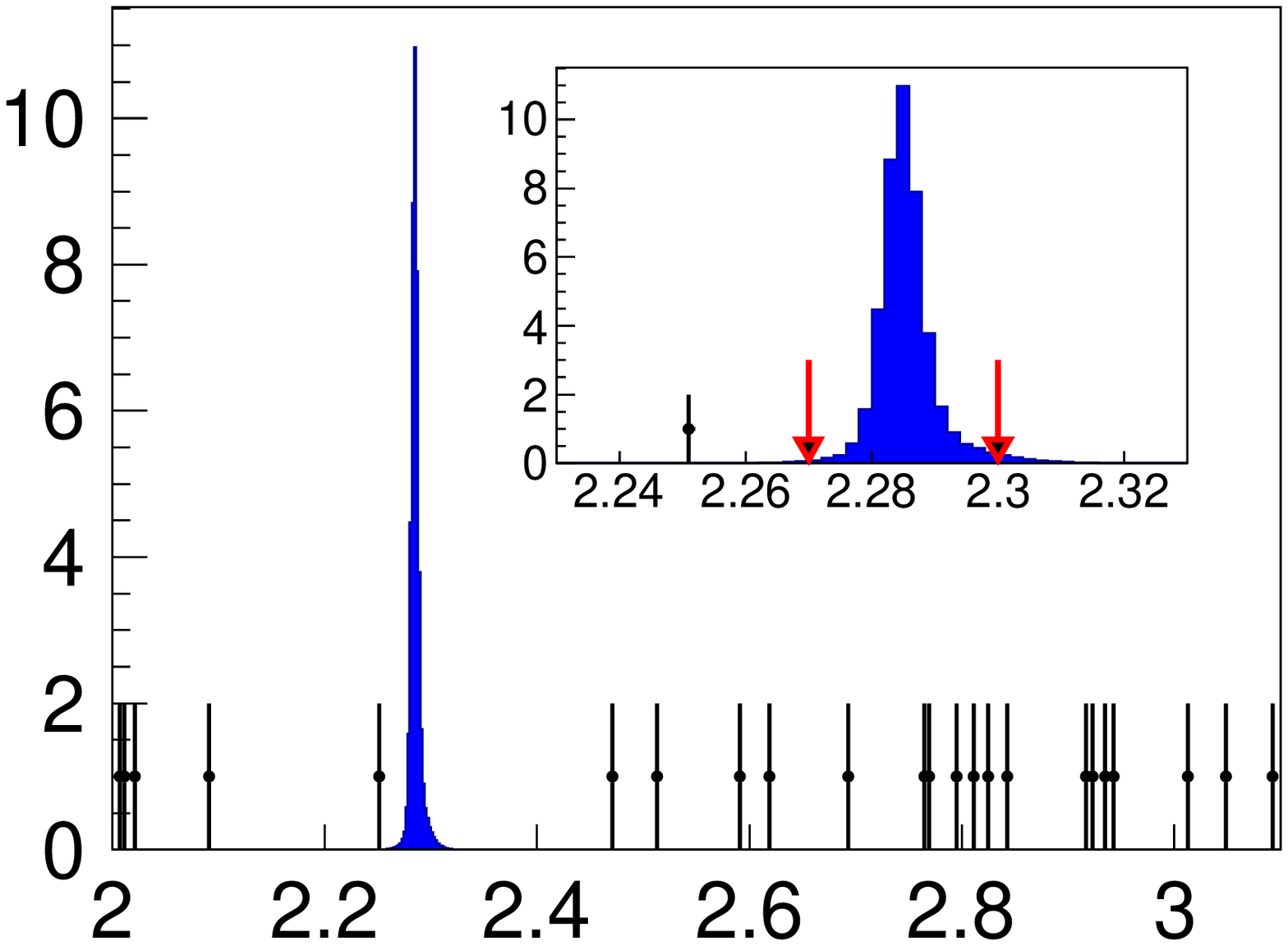}
\put(-235,25){\rotatebox{90}{\boldmath \bf\large Entries/($2$ MeV/$c^2$)} }
\put(-175,-10){\boldmath \bf\large $M_{p K^-\pi^+}$ (GeV/$c^2$)}
\caption{Distributions of $M_{pK^-\pi^+}$ for the $J/\psi\to\Lambda_c^+e^-$ candidate events for signal MC simulation (shaded histogram) and data (dots with error bars), where the signal MC sample is normalized arbitrarily. The inset plot shows a narrow mass range within (2.23, 2.33) GeV/$c^2$, where the arrows represent the signal mass window.}\label{fig::comparison-data}
\end{center}
\end{figure}

The candidate events of $J/\psi\to\Lambda_c^+ e^-$ are studied by examining the invariant mass of the $pK^-\pi^+$ system, $M_{pK^-\pi^+}$, as shown in Fig. \ref{fig::comparison-data}.
Since no events are observed in the signal window, the upper limit on the number of signal events $s_{90}$ for $J/\psi\to\Lambda_c^+e^-$ is estimated to be $5.7$ at the 90\% CL by utilizing a frequentist method \cite{ref::TROLKE} with unbounded profile likelihood treatment of systematic uncertainties, where the number of the signal and background events are assumed to follow a Poisson distribution, the detection efficiency is assumed to follow a Gaussian distribution, and the systematic uncertainty, which will be discussed below, is considered as the standard deviation of the efficiency. 
The upper limit on the branching fraction of $J/\psi\to\Lambda_c^+e^-$ is determined by
$$\mathcal{B}(J/\psi\to\Lambda_c^+e^-)<\frac{s_{90}}{N^{\rm tot}_{J/\psi}\times\mathcal{B}(\Lambda_c^+\to p K^-\pi^+)},$$
where $N^{\rm tot}_{J/\psi}=(1310.6\pm7.0)\times10^6$ is the total number of $J/\psi$ decays~\cite{ref::jpsi_num_inc},  and $\mathcal{B}(\Lambda_c^+\to pK^-\pi^+)= (6.35\pm0.33)\%$ is the decay branching fraction taken from Ref.~\cite{ref::pdg2016}.
Inserting the numbers of $s_{90}$, $N^{\rm tot}_{J/\psi}$ and ${\mathcal B}(\Lambda^+_c\to pK^-\pi^+)$ into the above equation, the upper limit on the branching fraction of $J/\psi\to\Lambda_c^+e^-$ is determined to be
$$\mathcal{B}(J/\psi\to\Lambda_c^+e^-) < 6.9\times 10^{-8}.$$

Systematic uncertainties in the measurement of $\mathcal{B}(J/\psi\to\Lambda_c^+e^-)$ mainly originate from the total number of $J/\psi$ events, the tracking efficiency, the PID efficiency, the kinematic fit, the MC modeling, and the quoted branching fraction for $\Lambda_c^+\to pK^-\pi^+$.
The uncertainty in the total number of $J/\psi$, determined via inclusive hadronic events, is $0.5$\%~\cite{ref::jpsi_num_inc}.
The uncertainty due to tracking efficiency is 1.0\% for each track, as determined from a study of the control samples $J/\psi\to pK^-\bar\Lambda$ and $\psi(3686)\to\pi^+\pi^- J/\psi$~\cite{ref::tracking}.
The uncertainties arising from the differences of PID efficiencies between data and MC simulation for electron, pion, kaon, and proton are determined with the control samples $e^+e^-\to\gamma e^+e^-$ (at $3.097$ GeV), $J/\psi\to K^+K^-\pi^0$, $J/\psi\to\pi^+\pi^-\pi^0$ and $J/\psi\to\pi^+\pi^-p\bar{p}$, respectively. They are $0.3$\%, $1.0$\%, $0.5$\% and $0.6$\% for electron, pion, kaon and proton, respectively.
The uncertainty of the kinematic fit is estimated using a control sample of $J/\psi\to\pi^+\pi^-p\bar p$, where a selection efficiency is defined by counting the number of events with and without the kinematic fit requirement. The difference of the selection efficiencies between data and MC simulation, $0.2$\%, is assigned as the corresponding systematic uncertainty.
The uncertainty due to MC modeling is negligible~\cite{ref::lambdac2pkpi}. 
In the calculation of the upper limit, the branching fraction $\mathcal{B}(\Lambda_c^+\to p K^-\pi^+)=(6.35\pm0.33)\%$ is quoted from Ref.~\cite{ref::pdg2016}, yielding a systematic uncertainty of $5.2$\%.
The total systematic uncertainty is $7.0$\%, obtained by adding all of the above uncertainties in quadrature.

In summary, by analyzing $1.3106\times10^9$ $J/\psi$ events collected at $\sqrt{s}=3.097$ GeV with the BESIII detector at the BEPCII collider, the decay of $J/\psi\to\Lambda_c^+e^-+c.c.$ has been investigated for the first time. No signal events have been observed and thus the upper limit on the branching fraction is set to be $6.9\times 10^{-8}$ at the $90$\% CL, which is more than two orders of magnitude more strict than that of CLEO's measurement in the analogous process~\cite{ref::cleo-d0}. The result is one of the best constraints from meson decays~\cite{ref::CLAS,ref::Babar} and is consistent with the conclusion drawn from the proton decay experiment~\cite{ref::SuperK}. 

The BESIII collaboration thanks the staff of BEPCII and the IHEP computing center for their strong support. This work is supported in part by National Key Basic Research Program of China under Contract No. 2015CB856700; National Natural Science Foundation of China (NSFC) under Contracts Nos. 11475090, 11575077, 11405046, 11335008, 11425524, 11625523, 11635010, 11735014; the Chinese Academy of Sciences (CAS) Large-Scale Scientific Facility Program; the CAS Center for Excellence in Particle Physics (CCEPP); Joint Large-Scale Scientific Facility Funds of the NSFC and CAS under Contracts Nos. U1532257, U1532258, U1732263; CAS Key Research Program of Frontier Sciences under Contracts Nos. QYZDJ-SSW-SLH003, QYZDJ-SSW-SLH040; 100 Talents Program of CAS; INPAC and Shanghai Key Laboratory for Particle Physics and Cosmology; German Research Foundation DFG under Contracts Nos. Collaborative Research Center CRC 1044, FOR 2359; Istituto Nazionale di Fisica Nucleare, Italy; Koninklijke Nederlandse Akademie van Wetenschappen (KNAW) under Contract No. 530-4CDP03; Ministry of Development of Turkey under Contract No. DPT2006K-120470; National Science and Technology fund; The Swedish Research Council; U. S. Department of Energy under Contracts Nos. DE-FG02-05ER41374, DE-SC-0010118, DE-SC-0010504, DE-SC-0012069; University of Groningen (RuG) and the Helmholtzzentrum fuer Schwerionenforschung GmbH (GSI), Darmstadt.


\end{document}